\documentclass[twocolumn,showpacs,preprintnumbers,amsmath,amssymb,prl,aps]{revtex4}
\usepackage{graphicx}
\usepackage{dcolumn}
\usepackage{bm}

\begin{document}

\title{Phase Separation of a Fast Rotating Boson-Fermion Mixture\\ 
in the Lowest-Landau-Level Regime}

\author{Rina Kanamoto
\footnote{Present address: Department of Physics, 
The University of Arizona, Tucson, AZ, USA.} and Makoto Tsubota}

\affiliation{
Department of Physics, 
Osaka City University, 
Osaka 558-8585, Japan}
\date{\today}

\begin{abstract}
By minimizing the coupled mean-field energy functionals, 
we investigate the ground-state properties of a rotating atomic 
boson-fermion mixture in a two-dimensional parabolic trap. 
At high angular frequencies in the mean-field-lowest-Landau-level regime, 
quantized vortices enter the bosonic condensate, 
and a finite number of degenerate fermions form 
the maximum-density-droplet state. 
As the boson-fermion coupling constant increases, 
the maximum density droplet develops into a lower-density state 
associated with the phase separation, revealing 
characteristics of a Landau-level structure.
\end{abstract}
\maketitle

%

Experimental developments in the cooling techniques of atomic 
gases have provided new opportunities to investigate 
the quantum-degenerate regime where Bose-Einstein condensations (BECs), 
Fermi-Dirac (FD) degeneracies~\cite{BFMex,DJ}, 
and BECs of paired states emerge~\cite{AtomMol,BCSBEC}. 
The heteronuclear Feshbach resonance, which was recently found 
in a boson-fermion mixture~\cite{Feshbach}, 
also enables us to investigate the quantum-statistical origin of novel 
phenomena. 
This system sheds light not only on our understanding of 
boson-mediated pairing of fermions~\cite{pairing}, but also on 
the relationship to condensed matter physics~\cite{CM}, 
multicomponent systems~\cite{Hamburg}, 
and the study of the normal and superfluid states 
of fermions~\cite{NS}. 

When a superfluid system is subjected to an external rotating drive 
above a critical angular frequency, the system forms quantized vortices, 
which are experimentally observed both in bosonic~\cite{BoseVor} and 
paired fermionic~\cite{FermiVor} systems. 
On the other hand, a number of analogies have been pointed out 
between neutral atoms subjected to an external rotation and 
charged particles in a magnetic field~\cite{HC,univ,WG} regardless of 
quantum statistics of atoms. 
As an example, a finite number of electrons in 
a quantum dot (QD) 
(e.g., realized in a semiconductor heterostructure) form 
the maximum density droplet (MDD) state~\cite{MDD}, 
reflecting the lowest-Landau-level (LLL) structure of electrons under 
a strong magnetic field. 
In the MDD, the total angular momentum of electrons takes 
the lowest possible value 
due to the FD degeneracy and the Pauli exclusion principle. 
However, the interplay between the Pauli pressure plus 
potential term and 
the electron-electron interaction causes an instability against 
density modulation or edge reconstruction~\cite{MDD-LDD,CW}. 
Such reconstructions are expected to emerge in the spin-polarized 
fermions under fast rotation. 


In this Letter, we address the boson-fermion mixture, and 
show that reconstruction process is associated with 
a phase separation between components due to the significance 
of repulsive boson-fermion interaction in the fast-rotating regime. 
For larger external angular frequency 
the components separate at smaller interaction, and 
the increase in the angular momentum of fermions 
from one of the MDD state is quantized at the integral multiple of the 
total number of fermions after phase separation. 
However, this quantization can be violated at the phase-coexistent regime 
as a consequence of the rotational symmetry breaking of the BEC with a vortex lattice. 

%

We consider the gaseous atomic mixture of $N_{\rm B}$ bosons and 
spin-polarized $N_{\rm F}$ fermions in a parabolic trap which 
rotates around the $z$ axis. 
When the angular frequencies of the trap are highly anisotropic 
as $\omega_z \gg \omega_{x,y}\equiv \omega$, 
the atomic motion virtually reduces to two-dimensional (2D). 
The trapping frequencies and atomic masses 
for the two components are assumed to be identical, being denoted as 
$\omega$ and $M$. 
Henceforth, the angular momenta, energies, and lengths 
are measured in units of 
$\hbar$, $\hbar\omega$, $l=\sqrt{\hbar/(M\omega)}$, respectively. 


The low-energy scatterings are characterized by $s$-wave scatterings 
which are modeled by contact interactions. 
The interaction and correlation between fermions are thus absent 
because of the Pauli principle. 
The dimensionless 2D boson-boson and boson-fermion couplings 
$g$, $h$ are related to the 3D ones 
$g_{\rm BB}^{\rm (3D)}=4\pi\hbar^2 a_{\rm BB}/M$ and 
$h_{\rm BF}^{\rm (3D)}=4\pi\hbar^2 a_{\rm BF}/M$, as 
$g=g_{\rm BB}^{\rm (3D)}/(\sqrt{2\pi}l_z)$ and 
$h=h_{\rm BF}^{\rm (3D)}/(\sqrt{2\pi}l_z)$ with 
$l_z=\sqrt{\hbar/(M\omega_z)}$, 
respectively.
We treat the boson-boson interaction and boson-fermion interaction 
within the mean-field approximation, which is valid for a weakly-interacting system 
under moderate rotating drive. As the rotating drive increases, 
both components will enter the {\it mean-field-lowest-Landau-level} 
(MF-LLL) regime~\cite{MFLLL}, 
where the single-particle states are described within the LLL, 
but the mean-field approximation is still valid~\cite{validity_condition}. 
Throughout this paper, we address the regime where the mean-field 
approximation is expected to be valid. 


Let us suppose that the bosons are BE-condensed occupying a 
single-particle state $\psi_{\rm B}$, while 
the fermions are FD-degenerate occupying single-particle orbitals 
$\psi_{\rm F}^{(j=1,...,N_{\rm F})}$. 
The mean-field energy functional in the rotating frame is given by
\begin{eqnarray}
E_{\rm B}&=&N_{\rm B}\int d^2r \psi_{\rm B}^*(\bm{r})
\left[\hat{H}+\frac{g}{2}n_{\rm B}(\bm{r})\right]
\psi_{\rm B}(\bm{r}),\\
E_{\rm F}&=&\sum_{j=1}^{N_{\rm F}}\varepsilon^{(j)}_{\rm F}; \quad 
\varepsilon^{(j)}_{\rm F}=\int d^2r \psi^{(j)*}_{\rm F}(\bm{r})
\hat{H}\psi_{\rm F}^{(j)}(\bm{r}),\\
E_{\rm BF}&=&h\int d^2r n_{\rm F}(\bm{r})n_{\rm B}(\bm{r})\label{EBF},
\end{eqnarray}
where $\hat{H}= (-\nabla^2+r^2)/2-\Omega\hat{L}_z$ 
is the Hamiltonian for a free atom with 
$\hat{L}_z=-i(x\partial_y-y\partial_x)$, and $\Omega$ is the 
angular frequency of the external rotating drive in unit of $\omega$. 
The densities are given by 
$n_{\rm B}(\bm{r})=N_{\rm B}|\psi_{\rm B}(\bm{r})|^2$, and 
$n_{\rm F}(\bm{r})=\sum_{j=1}^{N_{\rm F}}|\psi_{\rm F}^{(j)}(\bm{r})|^2$, 
where the summation over $j$ is taken over all occupied states of fermions. 
The eigensolutions of $\hat{H}$ 
\begin{eqnarray}
\varepsilon_{nm}^{(0)}\!\!\!&=&\!\!n+m+1-\Omega(m-n),
\quad n,m=0,1,...,\\
u_{nm}(\bm{r})\!\!\!&=&\!\!\frac{1}{\sqrt{\pi n!m!}}
e^{\frac{r^2}{2}}
\left(\partial_x+i\partial_y\right)^m
\left(\partial_x-i\partial_y\right)^n
e^{-r^2},\label{unm}
\end{eqnarray}
are well-defined angular-momentum states, i.e., 
$\hat{L}_zu_{nm}(\bm{r})=(m-n)u_{nm}(\bm{r})$. 


We expand the single-particle orbitals as 
$\psi_{\rm B}(\bm{r})=\sum_{n,m} b_{nm}u_{nm}(\bm{r})$, and 
$\psi_{\rm F}^{(j)}(\bm{r})=\sum_{n,m} f_{nm}^{(j)}u_{nm}(\bm{r})$, 
and minimize the total energy 
with the normalization conditions 
$\sum_{n,m} b_{nm}^2=\sum_{n,m} f_{nm}^{(j)2}=1$, 
where $b_{nm}$ and $f_{nm}$ are taken to be real without 
loss of generality. 
We then numerically iterate minimization of 
$\tilde{\varepsilon}_{\rm B}= (E_{\rm B}+E_{\rm BF})/N_{\rm B}$ 
and diagonalization of 
$\tilde{\varepsilon}_{\rm F}
=\varepsilon_{\rm F}+h\int d^2r \psi_{\rm F}^*n_{\rm B}\psi_{\rm F}$ 
until they self-consistently converges~\cite{BR,BFMF}. 
Since we use the mean-field approximation, 
we find a relevant breaking of rotational symmetry in 
the density distributions without investigating higher-order 
correlation functions. 

\begin{figure}[b]
\includegraphics[scale=0.47]{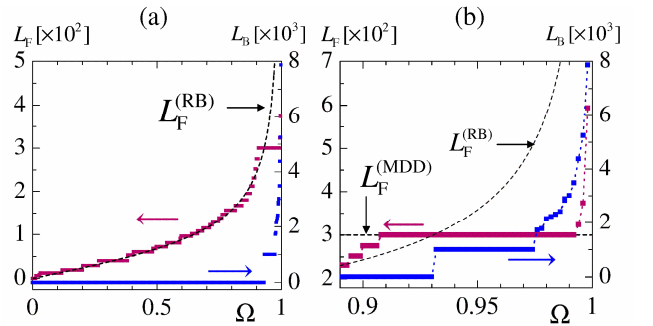}
\caption{(color online) 
Total angular momentum of each component with 
$N_{\rm B}=1000, N_{\rm F}=25$, and $g=0.5h=2\times 10^{-3}$. 
(a) Plateau at zero in $L_{\rm B}$ 
arises from the irrotationality. 
Small steps in $L_{\rm F}$ arise from the size effect. 
These steps become invisible for larger $N_{\rm F}(\gtrsim 500)$.
(b) Enlargement of (a) for the fast-rotating regime. 
}
\label{fig1}
\end{figure}


The ground-state angular momentum of each component is shown 
in Fig.~\ref{fig1} as a function of $\Omega$. 
The total angular momentum of the BEC, 
$L_{\rm B}=N_{\rm B}\sum_{n,m} b_{nm}^2(m-n)$, remains 
zero associated with the superfluidity. 
For fermions, in contrast, 
$L_{\rm F}=\sum_{j=1}^{N_{\rm F}}\sum_{n,m}f_{nm}^{(j) 2}(m-n)$ 
increases as $\Omega$ exceeds zero.
Within the semiclassical description with the rigid-body rotation, 
the total angular momentum of free fermions is given by 
$L_{\rm F}^{\rm (RB)}=\Omega\int r^2n_{\rm F}(\bm{r})d^2r
=(8N_{\rm F})^{3/2}\Omega/(24\sqrt{1-\Omega^2})$. 
This function is plotted with the dashed curve in Fig.~\ref{fig1}, 
and is found to agree well with $L_{\rm F}$ for slow rotating regime. 
The semiclassical theory of a free fermi gas is therefore 
good~\cite{BR_fermi} even when fermions weakly interact with bosons. 


As the angular frequency of the external rotating drive increases, 
$L_{\rm B}$ jumps from zero to $N_{\rm B}$ at
\begin{eqnarray}
\Omega_{\rm cr}^{\rm (B)}=1-G-2b_{11}^2-2Gb_{11}[b_{11}-(1+b_{00}^2)b_{00}]
\end{eqnarray}
where $G\equiv gN_{\rm B}/(8\pi)$, and $b_{11}^2=G^2/(8G^2+2G+1)$, 
$b_{00}^2=1-b_{11}^2$~\cite{derivation}. 
This jump corresponds to the onset of vortex formation in the BEC. 
The dominant contributions to $\Omega_{\rm cr}^{\rm (B)}$ are the first two terms 
which coincide with the result within the LLL approximation. 
The remaining terms arise from the higher Landau levels which are almost 
negligible as compared with the first two terms, but modify $\Omega_{\rm cr}^{\rm (B)}$ 
as $g$ increases. 
We find from the numerical and variational calculations 
that the BEC with $gN_{\rm B}\lesssim O(1)$ enters the MF-LLL regime 
in $\Omega \gtrsim \Omega_{\rm cr}^{\rm (B)}$ 
where the quantized vortices successively form in 
the BEC. 


For fermions the semiclassical description fails 
in a fast-rotating regime because the 
energy-level discreteness becomes crucial. 
This is shown in Fig.~\ref{fig1} (b), where the emergence of 
the plateau at $L_{\rm F}^{\rm (MDD)}$ and the significant 
deviation from $L_{\rm F}^{\rm (RB)}$ are the manifestations 
of the fact that fermions enter the LLL regime. 
The degenerate fermions occupy from 
the lowest level $m=0$ up to 
the highest level $m=N_{\rm F}-1$ in the LLL according to 
the Pauli principle, and hence $L_{\rm F}$ is frozen at the value 
$L^{\rm (MDD)}_{\rm F}=N_{\rm F}(N_{\rm F}-1)/2$ 
whereas $L_{\rm F}^{\rm (RB)}$ diverges. 
The order of $\Omega$ at which the fermions cross over from the semiclassical to 
a quantum regime is estimated by the condition 
$L_{\rm F}^{\rm (RB)}=L_{\rm F}^{\rm (MDD)}$, as
\begin{eqnarray}
\Omega_{\rm cr}^{\rm (F)}= \frac{3(N_{\rm F}-1)}{\sqrt{9N_{\rm F}^2+14N_{\rm F}-9}}, 
\end{eqnarray}
which approaches unity in the limit $N_{\rm F}\to \infty$. 


On the $L^{\rm (MDD)}_{\rm F}$-plateau the fermionic density becomes maximum, which 
is thus called a MDD in QDs with a finite number of electrons~\cite{MDD}. 
The area occupied by fermions becomes minimum since the mean-radius of 
the LLL orbital is given by $\langle r^2\rangle_m = m+1$. 
The MDD state is the trivial ground state of free fermions 
in the MF-LLL regime, and is regarded as 
the fermionic counterpart of a BEC without a quantized vortex in terms of 
the angular momentum~\cite{univ}. 
The corresponding many-body wave function of fermions 
is given by the Slater determinant constructed by 
the LLL orbitals $\psi_{\rm F}^{(j)}(\bm{r})=u_{0m}(\bm{r})$ for 
$m=0,1,\dots N_{\rm F}-1$, which is shown to be 
reduced to the Laughlin wave function of the integer quantum Hall state with $\nu=1$, 
\begin{eqnarray}\label{MDDwf}
\Psi_{\rm F}^{\rm (0)}({\cal Z}_1,{\cal Z}_2,\dots,{\cal Z}_{N_{\rm F}})
\!\!\!&=&\!\!\! \prod_{i<j}({\cal Z}_i-{\cal Z}_j)\exp\left(-\sum_{k}\frac{|{\cal Z}_k|^2}{2}\right),
\nonumber\\
\ &\ &\ \ \!\!\!\!\!
\end{eqnarray}
where ${\cal Z}\equiv x+i y$.
In the absence of the boson-fermion interaction, $L_{\rm F}$ 
remains $L_{\rm F}^{\rm (MDD)}$ for $\Omega \ge \Omega_{\rm cr}^{\rm (F)}$. 
In the presence of the boson-fermion interaction, however, 
$L_{\rm F}$ begins to increase in the limit $\Omega \to 1$ as shown in Fig.~\ref{fig1} (b). 
This behavior features a phase separation caused 
by the boson-fermion interaction, 
and is {\it not} the reminiscent of the divergence 
in $L_{\rm F}^{\rm (RB)}$ of free fermions. 
In the same manner as the nonlinear interaction of the BEC 
leads to the successive formation of quantized vortices, 
the boson-fermion interaction also induces the successive penetration of 
fluxes of angular momenta in the fermionic cloud. 


\begin{figure}[h]
\includegraphics[scale=0.47]{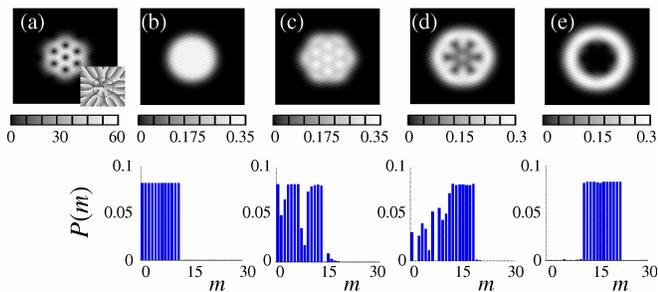}
\caption{(color online)
(a) Density, and phase (inset) profiles of the bosonic condensate 
wave function. 
(b)-(e) Density profiles (upper panels) and mean angular-momentum 
distributions (lower panels) of the fermionic cloud for 
$h/g=0,1,1.5,$ and $2.5$, respectively. 
}
\label{fig2}
\end{figure}

In order to examine how the boson-fermion coupling modulates 
the ground state in the MF-LLL regime, 
we henceforth restrict the bases $u_{nm}$ within the LLL, 
omitting the index $n$. 
Typical density profiles for several values of $h$ are shown 
in Fig.~\ref{fig2} where other parameters are fixed to be 
$\Omega=0.995, N_{\rm B}=1000, N_{\rm F}=12$, and $g=2\times 10^{-3}$. 
The number of quantized vortices remains unchanged, 
and hence the condensate wave function looks like ones shown in 
Fig.~\ref{fig2} (a) throughout the increase in $h$. 
In contrast, the fermionic density $n_{\rm F}$ changes drastically 
as $h$ increases. 
The histograms show the mean angular-momentum distribution 
of fermions defined by 
$P(m)\equiv \sum_{j={\rm occ}}|f_{m}^{(j)}|^2/N_{\rm F}$, 
which is equal to $1/N_{\rm F}$ for $0\le m \le N_{\rm F}-1$ 
in the MDD state [(b)]. 
When the weak interaction is introduced [(c)], the occupations of 
some low-angular-momentum states begin to be partially shifted to 
higher-angular-momentum states which are not occupied in 
the MDD state. The fermionic density increases in the vortex 
cores of the condensate, breaking the rotational symmetry 
in order to reduce the repulsive interaction energy. 
The derivation from the homogeneous distribution $1/N_{\rm F}$ in $P(m)$ 
is a signal of the rotational symmetry breaking. 
As $h$ increases further [(d)], more angular-momentum states 
are shifted, and the region where components overlap 
gradually decreases. The bosonic and fermionic densities 
eventually separate for larger values of $h$, where a large central 
hole emerges in the fermionic cloud [(e)]. 
The mean angular-momentum distribution again becomes uniform with the value 
$1/N_{\rm F}$ for $M\le m \le M+N_{\rm F}-1$, 
and the rotational symmetry in $n_{\rm F}$ 
is recovered independent of the configuration of the vortex lattice in the BEC. 
This state is regarded as the penetration of $M$ fluxes 
of angular momenta in the MDD state. Such a quasi-1D-like density and 
corresponding angular-momentum distribution 
also occur in the case of free fermions in $\Omega\gtrsim 1$ with 
an elimination of the centrifugal singularity~\cite{HC,AH}. 


\begin{figure}[b]
\includegraphics[scale=0.47]{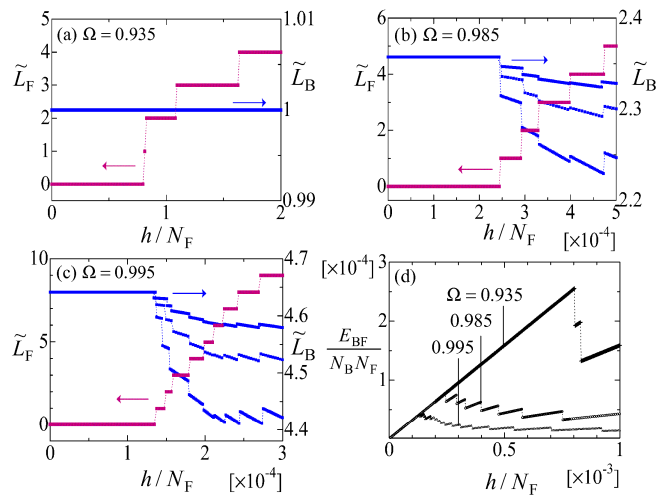}
\caption{
(color online) (a)-(c) Total angular momentum of each component 
for three values of $\Omega$. 
Three plots of $\tilde{L}_{\rm B}$ in each panel denote the results with $N_{\rm F}=32, 64, 128$ 
from top down. 
The plateaus in $\tilde{L}_{\rm F}$ are located at zero or positive integers. 
(d) Boson-Fermion interaction energy $E_{\rm BF}$. 
Jumps in $\tilde{L}_{\rm B}$, $\tilde{L}_{\rm F}$, and $E_{\rm BF}$ 
occur at the same values of $h/N_{\rm F}$ for a fixed $\Omega$. 
}
\label{fig3}
\end{figure}

We next study the changes in the total angular momentum of each component 
and the boson-fermion interaction energy associated with 
the phase separation in the MF-LLL regime. 
In Fig.~\ref{fig3} we show 
$\tilde{L}_{\rm B}\equiv L_{\rm B}/N_{\rm B}$, 
$\tilde{L}_{\rm F}\equiv (L_{\rm F}-L_{\rm F}^{\rm (MDD)})/N_{\rm F}$, and 
$E_{\rm BF}$ 
for several values of $\Omega$ and $N_{\rm F}$, with $g=2\times 10^{-3}$ and $N_{\rm B}=10^3$ 
being fixed. 
The changes in $\tilde{L}_{\rm B}$ depend on the size of fermions $N_{\rm F}$ 
while other quantities depend only on $h/N_{\rm F}$. 
Before the phase separation, 
$\tilde{L}_{\rm B}$ and $\tilde{L}_{\rm F}$ are nearly constant and 
$E_{\rm BF}$ linearly increases with $h/N_{\rm F}$. 
At a critical value $h/N_{\rm F}$ where $E_{\rm BF}$ becomes maximum, 
the phase separation occurs and fermions begin to rotate around the BEC. 
Since the boson-fermion interaction becomes significant for larger $\Omega$, 
the critical value of interaction $h$ becomes smaller for a faster rotating drive. 
After the phase separation, 
the system is well described in terms of 
{\it noninteracting} fermion model, 
because the fermions recover the rotational symmetry after phase separation 
and the value $P(m)$ cannot deviate from $1/N_{\rm F}$. 
The angular-momentum states of fermions in the phase-separating 
regime are therefore regarded as states where some angular-momentum fluxes 
penetrate the MDD. 
Let us define $M$ as the number of fluxes of angular momentum where 
the fermions homogeneously occupy the single-particle states ranging from 
$m=M$ to $m=M+N_{\rm F}-1$. 
The difference between the total angular momentum of fermions $L_{\rm F}^{(M)}$ 
with $M$ fluxes and one of the MDD state is given by
\begin{eqnarray}
L_{\rm F}^{(M)}-L_{\rm F}^{\rm (MDD)}=\sum_{m=M}^{M+N_{\rm F}-1}m-\sum_{m=0}^{N_{\rm F}-1}m=MN_{\rm F}. 
\end{eqnarray}
The corresponding many-body wave function of fermions with the angular momentum 
$L_{\rm F}^{(M)}$ is constructed 
by successive multiplications of the symmetric polynomial
\begin{eqnarray}
{\cal S}({\cal Z}_1,{\cal Z}_2,\dots,{\cal Z}_{N_{\rm F}})= \prod_j {\cal Z}_j
\end{eqnarray}
to the MDD state $\Psi_{\rm F}^{(0)}$ given by Eq.~(\ref{MDDwf}), 
where the antisymmetry of the fermionic wave function is ensured by $\Psi_{\rm F}^{(0)}$~\cite{MacDonald}. 
The polynomial ${\cal S}$ increases the angular momentum 
of each fermion by one, and the change in the total angular momentum $L_{\rm F}$ is thus 
given by $N_{\rm F}$. 
Upon the increase in $M$, the boson-fermion interaction energy $E_{\rm BF}$ decreases 
discontinuously as shown in Fig.~\ref{fig3} (d). The total angular momentum of bosons 
is partially transferred to fermions only when the number of quantized vortices in the BEC 
is larger than unity, and the number of vortices remains unchanged. 
On the other hand, the radius of fermionic cloud 
increases with $M$, while the rotational symmetry of the annular density profile 
is kept as shown in Fig.~\ref{fig2}(e). 

%

In conclusion, we have studied the phase separation between 
spin-polarized fermions and the BEC at high angular frequency 
in the MF-LLL regime. 
The ground-state reconstruction has been found 
in the density profiles, angular momenta, and boson-fermion interaction energy. 
We have shown that this reconstruction associated with the phase separation can be 
understood in terms of successive penetration of the angular-momentum fluxes to 
the fermionic cloud. 
We acknowledge Akira Oguri for fruitful discussion and comments. 




\begin{thebibliography}{88}
\bibitem{validity_condition}
For free bosons and free fermions of $N_{\rm B}=N_{\rm F}\equiv N$, 
the angular-momentum states are mapped as $L_{\rm F}=L_{\rm B}+N(N-1)/2$ each other. 
For weakly-interacting system, the MF-LLL regime for bosons corresponds to 
$O(N_{\rm B}) \lesssim L_{\rm B} \ll O(N_{\rm B}^2)$. 
This regime is mapped for fermions as 
$O(N_{\rm F}^2) \lesssim L_{\rm F} \ll O(N_{\rm F}^3)$, 
which is roughly expected to be a MF-LLL regime for fermions.
\bibitem{derivation}
This is derived with the assumption $h=0$. 
We have confirmed numerically that the boson-fermion 
interaction hardly affects $\Omega_{\rm cr}^{\rm (B)}$ 
while the boson-boson interaction is crucial.
\end{thebibliography}
\end{document}